\documentstyle[12pt,epsf,epsfig]{article}
\def\J{$J/\psi$}
\def\j{J/\psi}
\def\X{$\chi_c$}
\def\x{\chi}
\def\P{$\psi'$}

\def\C{c{\bar c}}

\def\be{\begin{equation}}
\def\ee{\end{equation}}

\def\lsim{\raise0.3ex\hbox{$<$\kern-0.75em\raise-1.1ex\hbox{$\sim$}}}
\def\gsim{\raise0.3ex\hbox{$>$\kern-0.75em\raise-1.1ex\hbox{$\sim$}}}

\def\NP{{ Nucl.\ Phys.\ }}
\def\PL{{ Phys.\ Lett.\ }}
\def\PR{{ Phys.\ Rev.\ }}

\def\PRL{{ Phys.\ Rev.\ Lett.\ }}

\def\ZP{{ Z.\ Phys.\ }}

\begin{document}

\thispagestyle{empty}

\noindent July 16, 2002 \hfill BI-TP 2002/14

\vskip 1.5 cm

\centerline{\Large{\bf Parton Percolation and \J~Suppression}}

\vskip 0.5cm

\centerline{\bf S.\ Digal, S.\ Fortunato, P.\ Petreczky and H.\ Satz}

\bigskip

\centerline{Fakult\"at f\"ur Physik, Universit\"at Bielefeld}
\par
\centerline{D-33501 Bielefeld, Germany}

\vskip 1cm

\centerline{\bf Abstract:}

\bigskip

The geometric clustering of partons in the transverse plane of nuclear
collisions leads for increasing $A$ or $\sqrt s$ to percolation.
In the resulting condensate, the partons are deconfined but not
yet in thermal equilibrium. We discuss quarkonium dissociation in this
precursor of the quark-gluon plasma, with an onset of dissociation when
the saturation scale of the parton condensate reaches that of the given
quarkonium state.

\vskip 1cm

Statistical QCD predicts color deconfinement for sufficiently hot
strongly interacting systems in full equilibrium. In the resulting
quark-gluon plasma, both the momenta and the relative abundances
of quarks, anti-quarks and gluons are determined by the tempe\-ra\-ture
of the medium. It is not evident if and at what evolution stage high
energy nuclear collisions produce such equilibrium, nor is it evident
that color deconfinement is restricted to such ideal thermal systems.
It thus seems natural to ask what conditions are necessary in the
pre-equilibrium stage to achieve deconfinement and perhaps subsequent
quark-gluon plasma formation. In recent years, the occurrence of color
deconfinement in nuclear collisions without assuming prior
equilibration has therefore been addressed on the basis of two closely
related concepts, parton percolation \cite{Armesto,Torino} and parton
saturation \cite{saturation,MV,reviewCGC}.

\medskip

Both start from the observation that in a central nucleus-nucleus
collision at high energy, one finds in the transverse nuclear
plane interacting
partons\footnote{A relation between deconfinement and percolation was
suggested quite long ago \cite{Baym}; but the first work on color
percolation in nuclear collisions was given in terms of strings
\cite{Armesto}, rather than partons. We shall nevertheless restrict
ourselves here to a parton picture \cite{Torino}.}
of different transverse scales.
At low densities,
one can define individual partons originating from nucleons of the
incident nuclei. Once the density of partons becomes so high that they
form a dense interacting cluster, independent parton existences and origins
are no longer meaningful: the resulting cluster forms a condensate
of deconfined partons. The condensate is formed in the sense of droplets
condensing to form a liquid, and the partons which make up this
condensate are no longer constrained by any hadronic conditions.

\medskip

Consider a distribution of partons of transverse size $\pi r^2$ over
the transverse nuclear plane $\pi R^2_A$, with $r << R_A$.\footnote{For 
simplicity, we assume the partons to have a fixed transverse radius; the 
extension to a distribution of radii is straight-forward and does
not change the picture.}
The fundamental aim of parton percolation studies \cite{Armesto,Torino,Baym}
is the determination of the transition 
from a normal hadronic collision situation of disjoint partonic `discs' to
a connected cluster of such discs 
spanning the nucleus, the parton condensate. Percolation theory predicts 
this transition
to occur when
\be
{N_q \over \pi R_A^2}~\pi r^2 = n_q~ \pi r^2 \equiv \eta \to
 \eta_c = 1.128
\label{1}
\ee
i.e., when the parton density $\eta = n_q \pi r^2$ measured in
terms of parton size reaches the percolation threshold $\eta_c$. In the 
`thermodynamic' limit of infinite spatial size
$R_A \to \infty$ and infinite parton number $N_q \to \infty$, the
largest connected cluster first spans the system at this point. 
The critical value
$\eta_c=1.128$ is determined by extensive numerical studies
\cite{perco}. Since the partons overlap, this does not mean that the
entire transverse nuclear surface is covered by parton discs. In fact,
at the percolation point, only the fraction $1-\exp(-\eta_c) \simeq 2/3$
of the nuclear area is covered by partons. At $\eta=\eta_c$, the
critical clustering behavior of the system can
be specified in the usual way in terms of critical exponents. In
particular, the size $S(\eta)$ of the largest cluster diverges for
$\eta \to \eta_c$ as
\be
S(\eta) \sim (\eta_c - \eta)^{-\gamma},
\label{2}
\ee
with the critical exponent $\gamma = 43/18$. While this holds
strictly only for infinite systems, it is verified that even for rather
small spatial systems, the transition from very small size to percolating
cluster occurs in a very narrow density interval. In other words,
even at finite size there is almost critical behavior. This will become
quite important for our further considerations.

\medskip

The essential idea of parton saturation is that the increase in the
number of partons for small $x$, as obtained from deep inelastic
scattering experiments, must stop when the density of partons becomes so
high that they overlap and form large interacting clusters; fusion and
splitting then causes their number to approach a constant. The onset of
saturation has been discussed in various ways; making use of their
transverse size, it can also be quite naturally determined by the
percolation condition (\ref{1}), which then fixes the saturation
scale in terms of $A$ and the c.m.s energy $\sqrt s$. Saturation
sets in when the parton density $n_q$ in
terms of the partonic interaction cross section $\sigma_q$ approaches
the critical value $\eta_c = 1.128$. The partonic cross section depends
on its inverse transverse momentum $k_T$, $\sigma_g(k_T) \sim 1/k_T^2$,
and the parton density $n_g(k_T^2)$ for a fixed resolution scale $k_T^2$
can be obtained from the gluon distribution function determined in deep
inelastic scattering. The novel aspect from the point of view of
percolation is that the density of partons $n_q(k_T^2)$ is related to
their transverse size $\sigma(k_T^2)$, so that with the functional form
of these two quantities given, the percolation condition specifies the
scale of the percolating partons. Let us consider this in more detail.

\medskip

The distribution of partons in an incident nucleon of momentum $p$ is
given in terms of their fractional momentum $x\simeq k/p$ and the
resolution scale $Q^2$, through integration over the partonic transverse
momentum $k_T$. The relevant resolution scale in a nucleon-nucleon
collisions is the largest transverse momentum $k_T^{\rm max}=Q$ for
which partons are resolved. The number of partons at fixed $x$ and
integrated over $k_T \leq Q$ is given by the sum of the contributions
of gluons plus those of sea quarks and antiquarks,
\be
{dN_q(x,Q^2)\over dy} = xg(x,Q^2) + \sum_i [xq_i(x,Q^2) +
x\bar q_i(x,Q^2)],
\label{3}
\ee
where $g(x,Q^2)$ denotes the gluon distribution function, $q_i(x,Q^2)$
and $\bar q_i(x,Q^2)$ that of up and down ($i=1,2$) quarks and antiquarks,
respectively. The distribution functions are determined from
parametrisations of deep inelastic scattering data, and thus eq.\
(\ref{3}) provides the number of partons at central rapidity $y=0$, with
$x=Q/\sqrt s$. We shall see shortly that for RHIC and higher energies,
the gluon contribution is strongly dominant; for SPS energy, however,
the quark and antiquark contributions cannot be neglected.

\medskip

We further need the parton size. In the simplest percolation approach,
which we shall follow here, this is just the geometric cross section
$\pi / Q^2$. More dynamical considerations lead to numerical
modifications, $\sigma_g(Q^2) = \kappa \alpha_s(Q^2) \pi / Q^2$, where
$\alpha_s(Q^2)$ is the running coupling at scale $Q^2$ and $\kappa$ a
given constant. With the geometric cross section we obtain that
in an $AA$ collision at $y=0$, the equation
\be
n_s(A) \left( {dN_q(x,Q^2) \over dy} \right)_{x=Q/\sqrt s}~
{\pi \over Q_s^2} = \eta_c,
\label{4}
\ee
determines the onset of percolation. Here $n_s(A)$ specifies the density
of parton sources in the transverse plane of a central $AA$ collision.
At SPS energy, this is essentially the density of wounded nucleons
\cite{NA50new}, $n_s(A) \simeq n_w(A)$. For higher energies and harder
partons, collision-dependent contributions will play a significant role,
and so a more suitable form here is a combination of the two sources,
\be
n_s(A) = (1-x)n_w + x n_{coll},
\label{5}
\ee
with $0 \leq x \leq 1$ \cite{KN}.

\medskip

As determined by eq.\ (\ref{4}), partons with $k_T \leq Q_s$ condense
to form an overlapping and hence interacting cluster spanning the
system, the parton condensate. Within this cluster, they can fuse or
split and thus lose their independent existence. We recall that at this
point, 2/3 of the nuclear area is covered by the parton condensate.

\medskip

The relation between saturation and percolation has so far not been much
emphasized. For a study of the new percolating medium at very large $A$
or $\sqrt s$, the `color glass condensate' of Ref.\ \cite{reviewCGC},
it is indeed not so important. However, it does become crucial for a
detailed picture of the onset of parton condensation. We know from
percolation theory that in the large volume limit this is a critical
phenomenon and hence even for finite systems takes place in an almost
singular way.

\medskip

To study the onset of parton percolation in $AA$ collisions, it is
convenient to rewrite eq.\ (\ref{4}) in the form
\be
{1\over Q^2}
\left( {dN_q(x,Q^2) \over dy} \right)_{x=Q/\sqrt s} = {25 \eta_c
\over \pi n_s(A)} \simeq {8.98 \over n_s(A)},
\label{6}
\ee
with the density of parton sources $n_s$ in fm$^{-2}$; the factor
25 arises when this is converted to GeV$^2$. In eq.\ (\ref{6}), the
hadronic parton distribution $(dN_q/dy)/Q^2$ is compared to the density
of parton sources in a nucleus-nucleus collision. For low source
densities, i.e., for small $A$, $8.98/n_s(A)$ remains well above the
$A$-independent l.h.s., which is a function only of $Q$ and $\sqrt s$:
there are not enough partons to form a condensate. For sufficiently
large $A$, however, $8.98/n_s(A)$ intersects $(dN_q/dy)/Q^2$ at some
$Q_s$, thus defining the onset of percolation. All partons with
$k_T\leq Q_s$ merge to form the condensate, in which interactions
prevent much further increase in the number of partons, i.e., there is
saturation.

\medskip

To see when that happens, we have to make use of a specific set of
parton distribution functions. The kinematic range relevant for our
analysis is 0.5 GeV$^2 < Q^2 < 2$ GeV$^2$, with $0.02 \le x <0.1$ for
SPS and $0.003<x<0.02$ for RHIC. Among the commonly used PDF
parametrizations, only the set GRV94 \cite{GRV94} goes down to such
small values of $Q^2$. This parametrization describes well the
available data on the proton structure function $F_2^p$ from the E665
and NMC collaborations for 0.4 GeV$^2 < Q^2 < 2$ GeV$^2$ and $x<0.01$
\cite{E665,NMC}. It also reproduces quite well the small $x$ HERA data
\cite{GRV94,GRV98} for $x>10^{-3}$. We have therefore calculated
$(dN_q/dy)$ using the next-to-leading order GRV94 PDF's in the DIS
scheme (GRV94DI) \cite{GRV94}; the resulting $(d N_q/dy)/Q^2$ are shown
in Figs.\ 1a and 1b for SPS (20 GeV) and RHIC (200 GeV) energies,
respectively.

\medskip

Any uncertainty in $(dN_q/dy)$ is mainly due to the gluon distribution.
At SPS energy, this can be estimated by comparing calculations using
leading and next-to-leading order GRV94 PDF's. The resulting
uncertainty is below 3 \% for $Q^2 > 0.6$ GeV$^2$ and increases to at
most 10 \% at 0.5 GeV$^2$. At RHIC, the next-to-leading order GRV94
PDF's are consistent with the gluon distributions determined by the H1
and ZEUS collaborations and with the constraints from charm
measurements \cite{Nagano}, while the leading order results
for the gluon distribution are too large. This gives us some confidence
in using the next-to-leading order GRV94 parametrization.

\medskip

We also note that these gluon distributions are not very different from
those proposed in more recent phenomenological saturation studies
for $x<0.01$ \cite{GBW}. This approach is very successful in describing
the low $x$ HERA data and it can in fact also account for the E665 data
on $F_2^p$ in the kinematical region relevant for RHIC. The gluon
distribution of \cite{GBW} is directly related to $F_2^p$; the sea quarks
are present only virtually as small dipoles from photon splitting
\cite{Mueller}. In such an approach, the uncertainty in the gluon
distribution can be avoided, and it has been used to predict the
$\sqrt{s}$-dependence of hadron multiplicities at RHIC \cite{KL}.

\medskip

Once the density of the percolating medium is sufficiently high, the
PDF approach of Ref.\ \cite{GBW} to the gluon distribution is more
appropriate to study the resulting condensate; however, it is not
suitable to study the onset of condensation. In this connection we note
that the effect of the sea quarks cannot be completely neglected even
at RHIC energy. In general, the relative weight of the different parton
species in the produced condensate will depend on $Q$ and $\sqrt s$,
and with increasing energy, the condensate becomes more and more
gluon-dominated. Thus the ratio of gluons to quarks and antiquarks in
the parton condensate based on the GRV94 PDF's is at $Q=1$ GeV found to
be about 1.2 for SPS energy and 4.0 for RHIC; in a chemically
equilibrated quark-gluon plasma, it is about 0.5. Hence before any
thermalization, the medium is strongly gluon-dominated \cite{Shuryak}.

\medskip

In order to illustrate the effect for different central $AA$ collisions,
we assume a spherical nuclear profile, which gives at SPS energy, with
$x=0$ in eq.\ (\ref{5}),
\be
n_s(A) = n_w(A) = {2A \over \pi A^{2/3}}.
\label{7}
\ee
Inserting this into eq.\ (\ref{6}), together with the values of
$[dN_q(x,Q^2)/dy]/Q^2$ at $\sqrt s = 20$ GeV of the previous section,
we obtain the results shown in Fig.\ 1a. The intersection points
determine the onset of percolation. It is seen that in this very
simplified picture, parton condensation begins for $A \simeq 60$, with
$Q \simeq 0.7$ GeV. To obtain the corresponding behavior at $\sqrt s =
200$ GeV, nucleon collisions have to be included as source of partons.
With
\be
n_{coll}={3 \over 4} \left( {A^{4/3} \over \pi A^{2/3}} \right);
\label{8}
\ee
and $x=0.09$ \cite{KN} in eq.\ (\ref{5}), we find the percolation
points in Fig.\ 1b. In eq.\ (\ref{8}), the factor 3/4 comes from
averaging over the nuclear profile. In Fig.\ 2, the percolation values
$Q_s$ for central $AA$ collisions are displayed as function of $A$. At
SPS energy, we thus do not obtain parton condensation below $A \simeq
60$; at RHIC energy, the higher parton density lowers the onset to
$A \simeq 40$.

\medskip

In the experimental study of \J~suppression, the production is measured
at different centralities, so that we now have to determine the
parton source density at fixed $A$ and varying impact parameter.
This is done in a Glauber analysis based on Woods-Saxon nuclear
profiles, with a collision-determined weight \cite{KLNS}. In Fig.\ 3a,
we show the resulting percolation behavior as function of the
effective number $N_{\rm part}$ of participants in a $Pb-Pb$ collision
at $\sqrt s =20$ GeV. The threshold for parton percolation is found to
be slightly below $N_{\rm part} \simeq 150$. The corresponding
calculations for $Au-Au$ collisions at $\sqrt s = 200$ GeV, with a
collision-dependent term in $n_s$ and again $x=0.09$ in eq.\ (\ref{5}),
lead to the results shown in Fig.\ 3b. The onset of parton condensation
at RHIC is thus shifted to considerably more peripheral collisions. In
Fig.\ 4, the centrality dependence of the percolation scale $Q_s$ is
shown; at the onset point, the condensate contains partons of different
sizes $r$, with $r \geq 1/(0.7~ {\rm GeV})$ at SPS and $r \geq
1/(0.9~{\rm GeV})$ at RHIC.

\medskip

We now turn to the effect of parton condensation on \J~production.
It is known from $pA$ collisions that normal nuclear matter leads to
reduced charmonium production. Therefore we first have to consider such
`normal' suppression, since for $pA$ collisions, at least up to RHIC
energies, we are well below the threshold for parton
condensation\footnote{A determination
of the collision energy for which parton saturation occurs in $pp/p
\bar p$ or $pA$ collisions is presently difficult, since it involves
$x$-values much smaller than presently accessible in deep inelastic
scattering. Disregarding this small $x$-behavior leads to a very early
onset of parton percolation \cite{Dias}.}.
Consider the production of charmonium in the nuclear target rest
frame. A gluon from the incident proton fluctuates into a virtual
$\C$ pair; in a collision with one of the target nucleons this is
brought on-shell, its color is neutralized, and it eventually becomes
a physical \J~of size $d_{\j}\simeq 0.4$ fm. Thus in the rest-frame of the
$\C$ pair, a time of at least 0.4 fm is needed for \J~formation. During
its pre-resonance stage, for $\tau_{\C} \leq 0.4$ fm, it will travel a
distance
\be
d_{\C}= \tau_{\C} {sx_F \over 4Mm}\left[1 + \sqrt{1+ {4M^2 \over sx_F^2}}
\right]
\label{9}
\ee
in the target rest frame; here $m$ denotes the mass of the nucleon,
$M$ that of the $\C$ pair and $x_F$ the $\C$ Feynman momentum
fraction. For the production of a $\C$ at rest in the nucleon-nucleon
c.m.s., this becomes $d_{\C} = \tau_{\C}(\sqrt s/ 2m)$. From this it
is immediately seen that at $\sqrt s \simeq 40$, and $x_F \geq 0$,
the nascent \J~effectively traverses even heavy nuclear targets in
its pre-resonance stage. The situation is very similar for $\x_c$
and \P~production, for which the pre-resonance life-times are if
anything even larger. Thus for the mentioned $\sqrt s$ and $x_F$,
the nuclear target sees of all charmonium states only the small
pre-resonance precursor, so that all should suffer the same degree of
nuclear suppression \cite{KS96}. For negative $x_F$ (and perhaps to some
extent also for lower $\sqrt s$), the nucleus should begin to see
physical resonances, and as a result the suppression should become
larger for the higher excited states with their larger radii \cite{KS95}.

\medskip

On the basis of this information, normal charmonium suppression has
generally been studied in terms of pre-resonance dissociation in
standard nuclear matter, leading to a break-up cross section around
5 - 6 mb \cite{KLNS,Shahoyan}. This implies a mean free path of about
12 fm. For a multiple collision analysis of Glauber type, the mean free
path has to exceed the coherence length (essentially the size) of the
pre-resonance in the rest frame of the nucleus: otherwise, the $\C$ is
wounded several times before it has had a chance to register the first
interaction. For collision energies $\sqrt s \leq 40$ GeV, i.e., in the
range of fixed target experiments, this condition is satisfied, with
coherence lengths below 8 fm. At RHIC energy, on the one hand the cross
section for pre-resonance break-up in normal nuclear matter could
change, and on the other hand the coherence length becomes dilated ten
times more. Now interference effects of the
Landau-Pomeranchuk-Migdal type may have to be taken into account, which
could lead to a reduction of `normal' nuclear suppression \cite{Fujii}.
Hence measurements of charmonium production in $pA$ collisions at
RHIC are absolutely essential for an understanding of whatever
\J~suppression in $AA$ collisions is observed there.

\medskip

We now want to consider the additional `anomalous' suppression of
charmonium production due to the dense partonic medium created in $AA$
collisions. The virtual partons in the incoming nuclei coalesce to form
a condensate in a time $t_c \simeq 1/Q_s$ determined by the
saturation scale; in the color glass picture \cite{reviewCGC}, this is
the time needed to melt the frozen glass. The interacting and expanding
parton condensate can subsequently lead to the formation of a
thermalized quark-gluon plasma; a crucial factor for this is the energy
density reached at thermalization. We restrict ourselves here to the
parton condensate stage.

\medskip

In $AA$ collisions at RHIC energy, the colliding nuclei are in the
overall c.m.s Lorentz-contracted to about 0.1 fm. They will therefore
sweep past the nascent charmonium in its pre-resonance state and
before parton condensation sets in, resulting in some form of normal
nuclear absorption. After about 0.2 fm, the nuclei are well out of
the way and the parton condensate is formed. Any produced and surviving
charmonium states now encounter this new medium, either as fully formed
resonances or in the late pre-resonance stage. For SPS energy, a
similar discussion is more complex, since the nuclei are contracted to
only about 1 fm diameter; this will introduce a smearing in the
comparison of the different time scales. We shall here neglect this and
assume that also at the SPS there is first pre-resonance absorption in
normal nuclear matter, followed by the effect of the parton condensate
on the survivors.

\medskip

In the first attempt to describe \J~suppression in terms of parton
percolation \cite{Torino}, it was assumed that different charmonium
states $i$ define particular scales $Q_i$, and the onset of percolation
for partons of that scale leads to the dissociation of all
charmonia of that species within the percolating cluster. The number of
partons was taken as a scale-independent function of $\sqrt s$. Parton
saturation in fact specifies $[dN_q/dy](Q_s^2)$ as function of the scale
$Q_s$, so that we now have (see Figs.\ 2 and 4) at given collision
energy a parameter-free determination of the onset line $Q_s(A)$ or
$Q_s(N_{part})$ of parton condensation.

\medskip

Within the parton condensate, color fields of a $Q_s$-dependent strength
will affect the binding of a $\C$ dipole 
charmonium state. This effect can be addressed in different ways.
One possible way is to consider the $\C$ propagation in a classical
field, which can represent the gluon field of the condensate or that of
the nucleus; the information on specific medium is encoded in the
corresponding field correlator \cite{Fujii}. For a small singlet dipole
the probability to remain singlet is proportional to
$\exp(-{\kappa} r^2)$, where $r$ is the dipole size and $\kappa$ is a
dimensional parameter determined by the field strength. In the color
glass approach, $\kappa \sim Q_s^2$ \cite{reviewCGC}. Thus for small
singlet dipoles, $Q_s^2 r^2 << 1$, the probability to remain in a singlet
state is close to unity (color transparency). Motivated by this fact we
shall here adopt the simple model of Ref.\ \cite{Torino}, assuming
that a charmonium state $i$ of scale $Q_i$ will be dissociated if it
finds itself in a parton condensate of scale $Q_s \geq Q_i$; otherwise
it will survive. This very simplistic picture allows an analysis of
nuclear profile effects (condensed and non-condensed regions of the
collision profile) and thus provides some direct predictions for
the centrality-dependence of anomalous \J~suppression for given
$A$ and $\sqrt s$. Other approaches that have been suggested include
the study of the time evolution of the screening masses in the parton
cascade \cite{Dinesh} and in the color glass condensate \cite{Goncalves}.

\medskip

The radii of the observable charmonium states as obtained from
the solution of the Schr\"odinger equation with Cornell potential
are \cite{FK}
\be
r_{\j}\simeq (0.9~{\rm GeV})^{-1},~
r_{\xi}\simeq (0.6~{\rm GeV})^{-1},~
r_{\psi'}\simeq (0.45~{\rm GeV})^{-1}
\label{10}
\ee
These have to be compared to the partonic saturation radii $Q_s^{-1}$
shown in Fig.\ 4 as functions of the number of participants in central
$AA$ collisions. It is seen that at SPS energy, the onset of \X~and
\P~suppression effectively coincides with the onset of parton
condensation, while the \J~survives up to larger $Q_s$ and higher
parton densities\footnote{The fate of the \P~is most likely more
complex. It should certainly be dissociated once parton condensation
sets in; however, it is very much more weakly bound than \X~and \J, so
that a less dense environment could also lead to its break-up.}. Hence
in $Pb-Pb$ collisions, anomalous suppression should start with the
elimination of \J~feed-down from \X~and \P~at $N_{part} \simeq 150$; the
further suppression of direct \J~production should set in for $N_{part}
\simeq 250$. In Fig.\ 5 we show the suppression pattern observed by the
NA50 collaboration at the CERN-SPS \cite{NA38/50}; the two `steps'
observed in the anomalous suppression pattern agree fairly well with the
expected onsets of parton condensation. Note that at the percolation
points, the percolating cluster covers only a fraction of the transverse
area; with increasing centrality, this fraction increases. Hence more
detailed studies are needed to determine the actual amount of
suppression as function of centrality (see \cite{Torino}); the present
study only indicates the onset points.

\medskip

Finally we want to consider the suppression pattern expected for RHIC
experiments. From Fig.\ 4, it is seen that for $N_{part} \geq 80$, all
charmonium states should suffer anomalous suppression, so that here
there should only be one onset point. Moreover, collisions with
$N_{part}=80$ correspond to an impact parameter of $b \simeq 10$ fm, and
this may be too peripheral for meaningful measurements. For a study of
the onset of anomalous suppression at RHIC, it will thus most likely be
necessary to study $AA$ collisions for much lower $A$, as already noted
previously \cite{Torino}.

\medskip

In conclusion, we have shown that a very simplistic parton percolation
approach leads to a conceptually reasonable and effectively parameter-free 
description of the observed anomalous $J/\psi$ suppression pattern.

\bigskip

\centerline{\bf Acknowledgements}

\bigskip

It is a pleasure to thank D.\ Kharzeev and M.\ Nardi for their
contribution to an early version of this study \cite{Torino} as well as
for many helpful discussions and comments.


\newpage
\begin{figure}
\centerline{\epsfxsize=13cm \epsffile{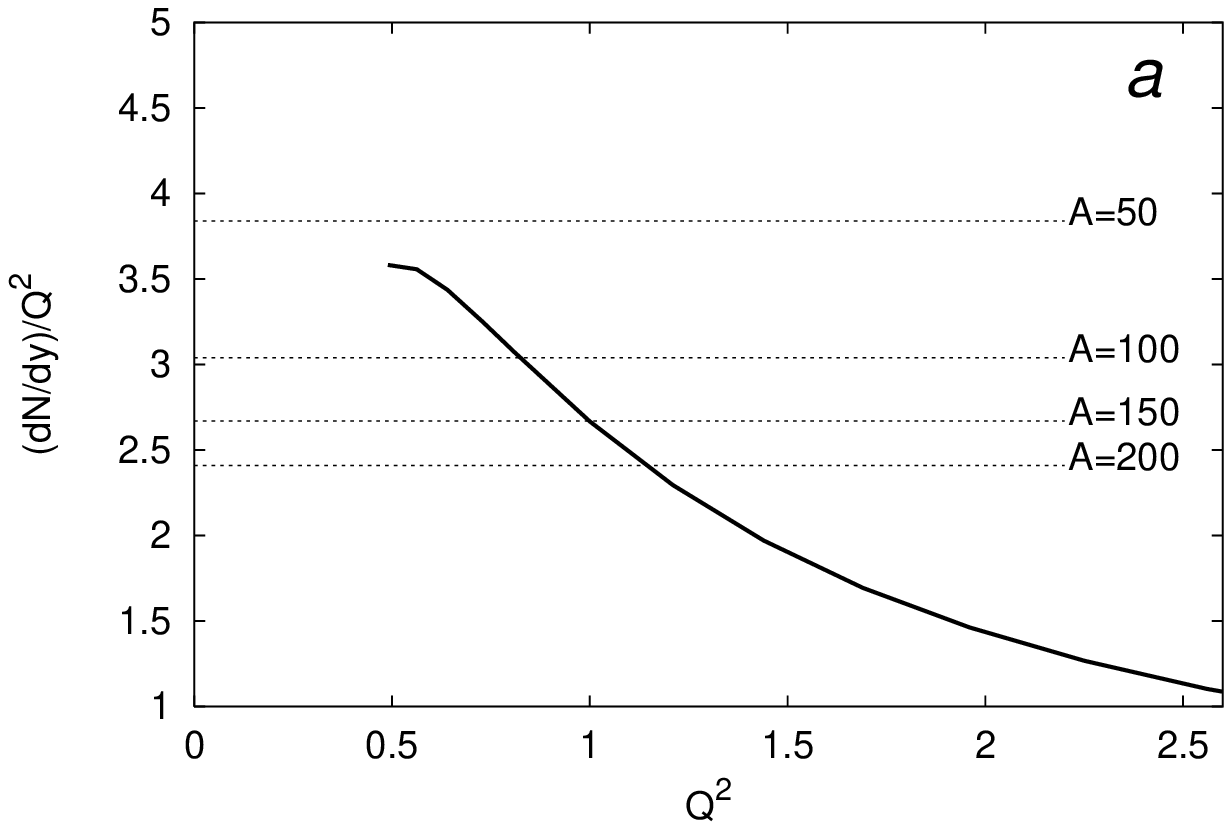}}
\centerline{\epsfxsize=13cm \epsffile{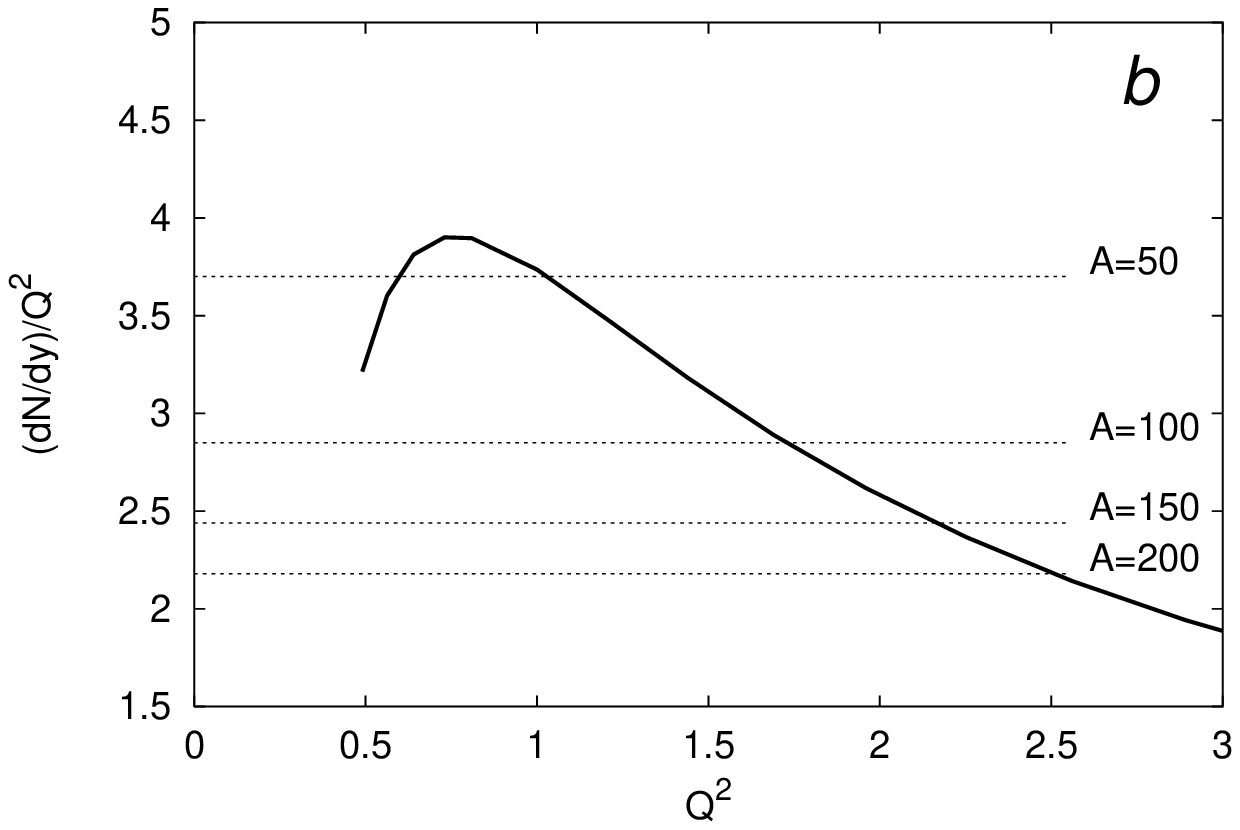}}
\caption{
Parton distribution vs.\ $Q^2$, with the percolation limits
obtained for different central $AA$ collisions: (a) $\sqrt s = 20$ GeV,
(b) $\sqrt s = 200$ GeV.}
\end{figure}

\begin{figure}
\centerline{\epsfxsize=13cm \epsffile{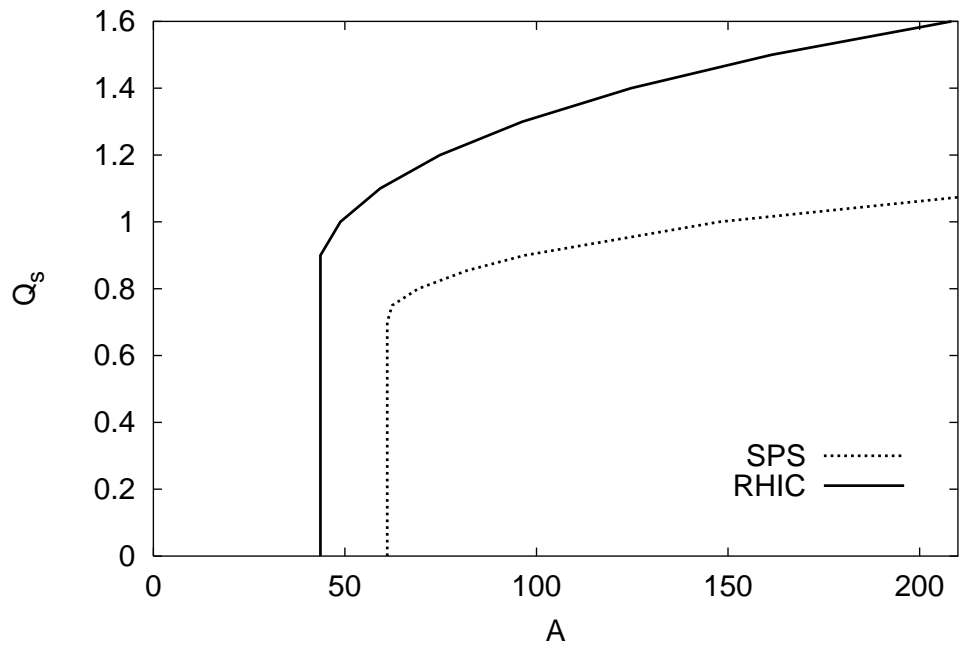}}
\caption{Saturation momenta $Q_s$ vs. $AA$ for central $AA$ collisions at
SPS and RHIC energies.}
\end{figure}

\begin{figure}
\centerline{\epsfxsize=13cm \epsffile{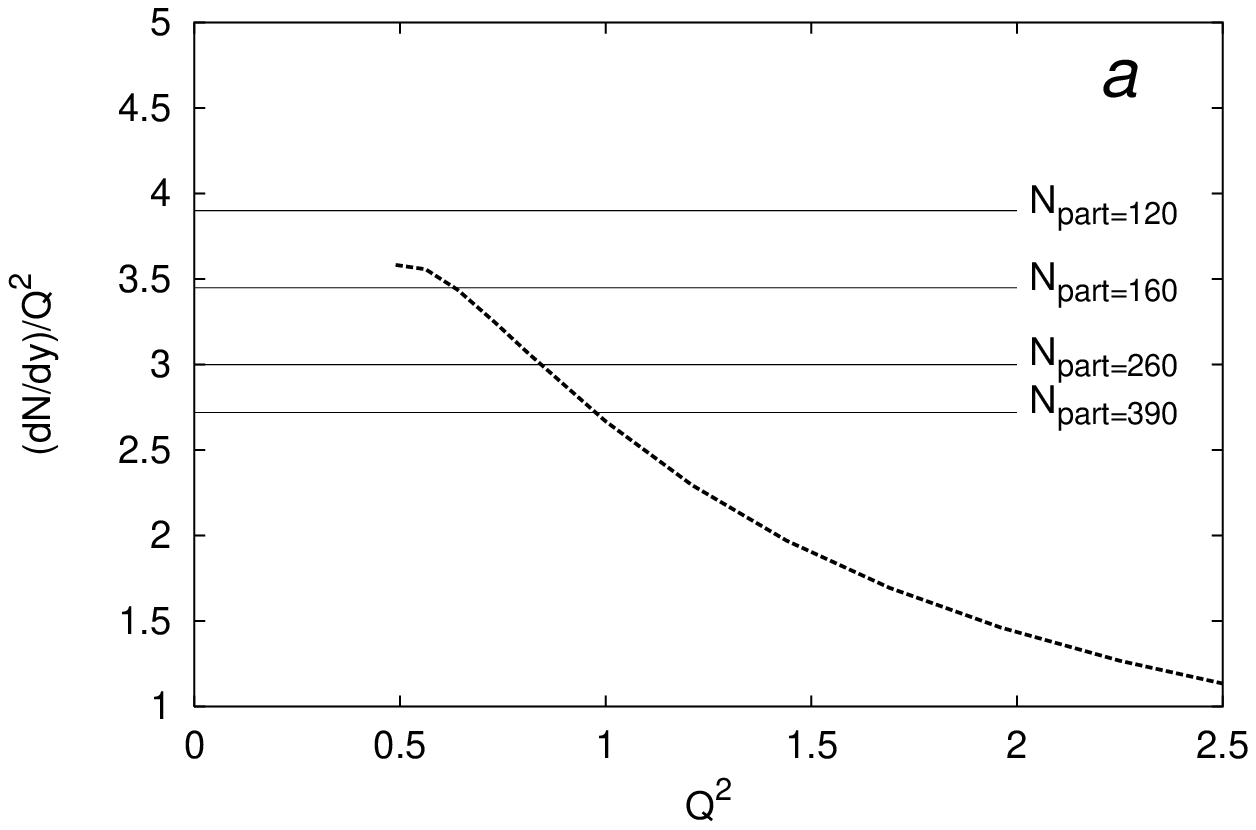}}
\centerline{\epsfxsize=13cm \epsffile{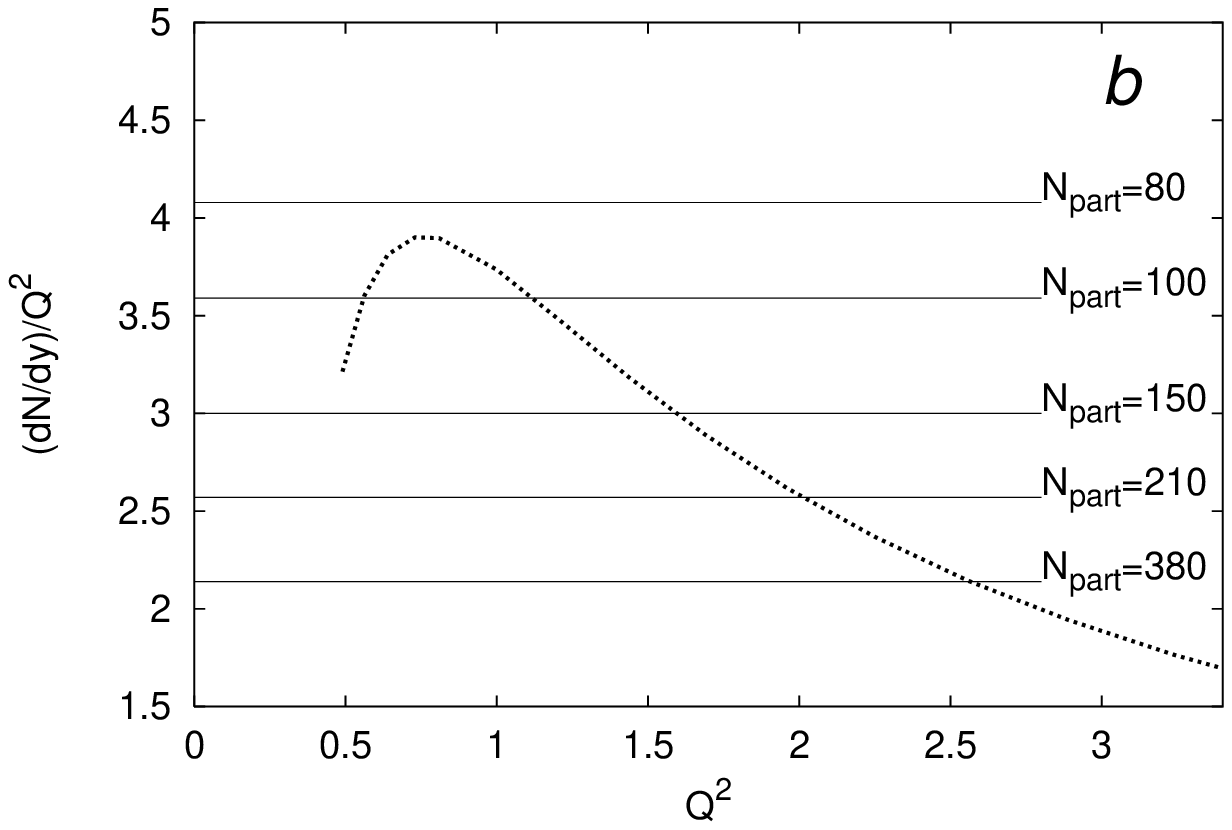}}
\caption{
Parton distribution vs.\ $Q^2$ with the percolation limits 
for different centralities, given by the number of participant nucleons;
: (a) $Pb-Pb$ at $\sqrt s = 20$ GeV,
(b) $Au-Au$ at $\sqrt s = 200$ GeV.
}
\end{figure}

\begin{figure}
\centerline{\epsfxsize=13cm \epsffile{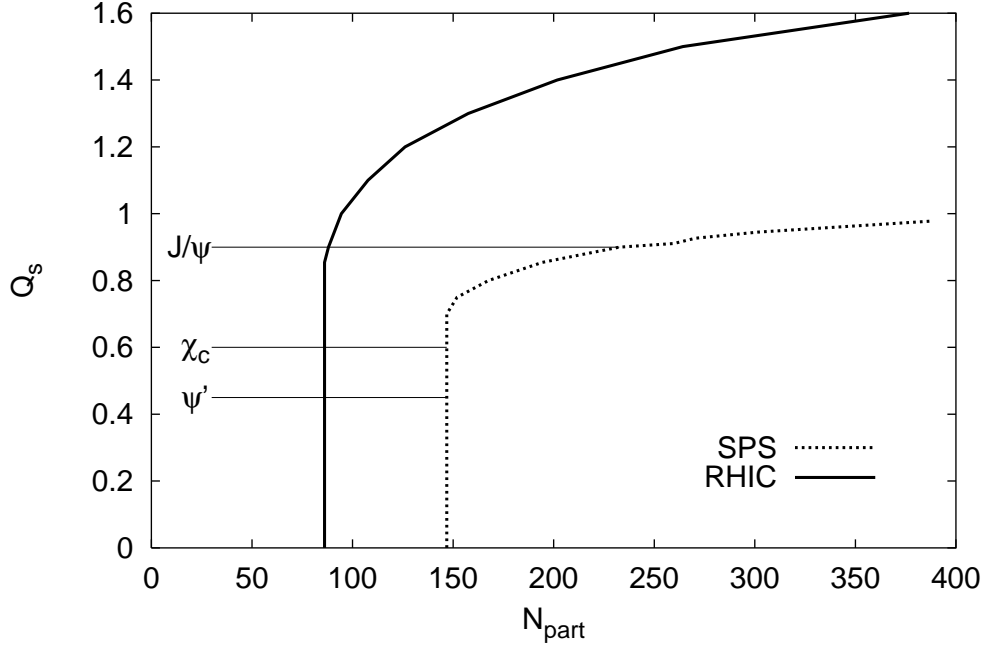}}
\caption{Saturation momenta $Q_s$ vs. $AA$ for central $Pb-Pb$ collisions
at $\sqrt s = 20$ GeV and central $Au-Au$ collisions at $\sqrt s =200$ GeV,
together with the critical scales for anomalous \J, \X~and \P~suppression.
}
\end{figure}

\begin{figure}
\centerline{\epsfxsize=13cm \epsffile{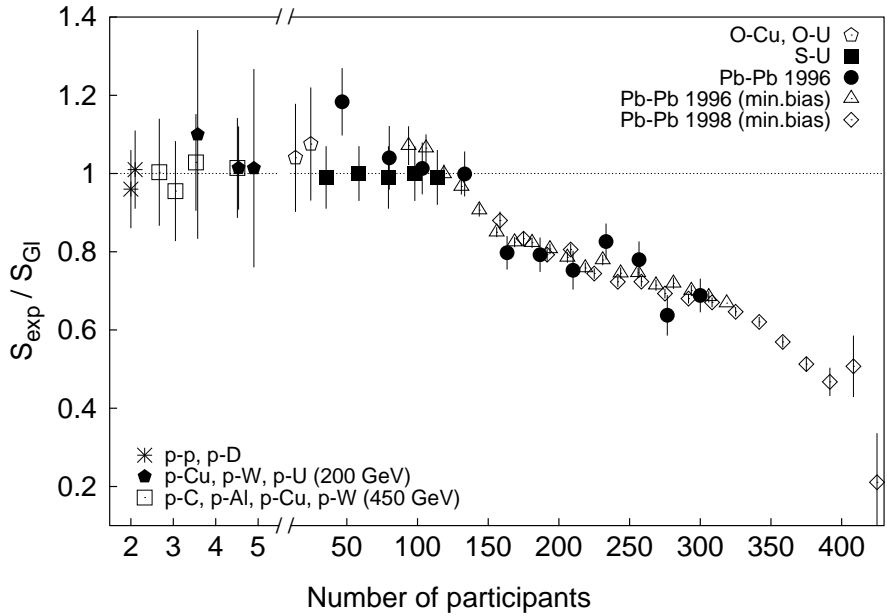}}
\caption{
The \J~survival probability determined at SPS energy as function
of the number of participant nucleons, from \cite{NA38/50}.
}
\end{figure}

\end{document}